\documentclass[reprint,longbibliography]{revtex4-2}%
\usepackage{amssymb}
\usepackage{graphicx}
\usepackage{amsmath}
\usepackage{amsfonts}%
\providecommand{\U}[1]{\protect\rule{.1in}{.1in}}
\begin{document}

\title{Breakup to non-breakup transition of air entrained into viscous liquid by a
disk: analogy of the self-similar dynamics with critical phenomena}
\author{Shoko Ii}
\author{Ko Okumura}
\email{okumura@phys.ocha.ac.jp}
\affiliation{Physics Department and Soft Matter Center, Ochanomizu University, 2-1-1\\
Ohtsuka, Bunkyo-ku, Tokyo 112-8610, Japan}
\date{\today}

\begin{abstract}
Self-similarity in partial differential equations has been widely exploited to
study many phenomena in physical sciences. We have studied the interfacial
dynamics when air is entrained into viscous liquid by a disk in a confined
geometry. In a previous study using an original experimental system, we found
the sheet- and corn-forming regimes, in which a sheet and cone of air are
respectively formed before air detaches from the disk. The sheet eventually
breaks up but the corn, which appears when a bit more confined, does not.
Here, we find a third regime, in which a corn eventually breaks up, by
investigating different ranges of confining parameters: the transition from
breakup to non-breakup can occur within the corn regime. Furthermore, with the
data obtained in the third regime we deeply explore analogy with critical
phenomena to find out that the counterpart of the critical exponents dependent
on a length scale. Since the scale is a number not discrete but continuous,
the present hydrodynamic analog suggests the existence of an uncountably
infinite number of universality classes. The rich physics revealed in our
study suggests a promising direction of the study of the self-similar
dynamics: exploring analogy with critical phenomena, focusing on confined
geometries in many natural and industrial phenomena.

\end{abstract}
\maketitle

\affiliation{Physics Department and Soft Matter Center, Ochanomizu University, 2-1-1\\
Ohtsuka, Bunkyo-ku, Tokyo 112-8610, Japan}

\affiliation{Physics Department and Soft Matter Center, Ochanomizu University, 2-1-1\\
Ohtsuka, Bunkyo-ku, Tokyo 112-8610, Japan}

\affiliation{Physics Department and Soft Matter Center, Ochanomizu University, 2-1-1\\
Ohtsuka, Bunkyo-ku, Tokyo 112-8610, Japan}

\affiliation{Physics Department and Soft Matter Center, Ochanomizu University, 2-1-1\\
Ohtsuka, Bunkyo-ku, Tokyo 112-8610, Japan}

\section{Introduction}

When shapes as a function of $x$ defined by a solution $h(t,x)$ of a partial
differential equations (PDE) at different $t$ collapse onto a single master
curve, the dynamics is called self-similar. The self-similarity in PDEs has
played a vital role in understanding many phenomena in physical sciences
\cite{Barenblatt1979,barenblatt2003scaling}, in particular, phenomena
categorized as the singular dynamics
\cite{kadanoff1997singularities,eggers2015singularities}, which include
fluid-jet formation
\cite{2000NatureLathropFluidJetEruption,2002PRLCohenSelectiveWithdrawal}, drop
coalescence \cite{YokotaPNAS2011,hernandez2012symmetric,kaneelil2022three},
gravitational collapse of stars \cite{choptuik1993universality} and cell
aggregation \cite{Cellaggregation1995}. Among them, the formation of a fluid
droplet occurring, for example, in a dripping faucet, has been extensively
studied. For the breakup of a liquid drop in air, a dynamical regime, in which
inertia, capillary, and viscosity compete among others, is known
\cite{1993PRLEggersPinchoff,1994ScienceNagelDropFallingFaucet}, while, for the
breakup of a bubble, a regime, in which high viscosity of the surrounding
fluid competes with capillarity, is found
\cite{2003ScienceNagelMemoryDropBreakup}. The latter regime is further found
to exhibit a crossover to another viscous-capillary regime in the breakup of a
bubble in a tube \cite{pahlavan2019restoring}. In all these examples, the
self-similar dynamics of the breakup of a fluid drop is axisymmetric.

Recently, however, an example of the self-similar drop breakup without
axisymmetry was reported \cite{nakazato2018self} in a confined geometry. In
the experiment, we developed an original experimental system, in which a metal
disk entrains air into viscous liquid in a confined geometry, to find
\textit{the sheet-forming regime with breakup}, in which the entrained air
eventually detaches from the disk \textit{with forming a bubble}: before the
detachment a non-axisymmetric constriction region appears in the interface
where \textit{a sheet of air} is formed, which thins down to breakup. In the
second study, by using the same original experimental system, we further
revealed another regime, \textit{the corn-forming regime without breakup}. In
this regime, a bubble is \textit{not} created: the constriction point does not
appear in the interface, but \textit{a corn of air}, which is actually
non-axisymmetric, is formed instead of a sheet near the detachment point, as a
result of further confining the system \cite{nakazato2022air}. (We remark here
that these regimes were found in the pre-detachment dynamics, while the
post-breakup dynamics in the sheet-forming regime has recently been studied
\cite{Yoshino}.) Here, we find a third regime, the corn-forming regime
\textit{with breakup}, in the pre-detachment dynamics, in which the
constriction region appears and then a bubble is created but a corn of air,
instead of a sheet, is formed at the breaking tip, by investigating different
ranges of confining parameters. In other words, we find that the transition
from breakup to non-breakup can occur within the corn-forming regime, while in
any cases the constriction region appears for a breakup but does not for
non-breakup: the appearance of the constriction point is a sign of the breakup.

We further explore, in the third regime, analogy with critical phenomena in
thermodynamic transitions \cite{Cardy,Goldenfeld}. We show that the master
curve, or \textit{scaling function} for the breakup case is distinctively
different from that for the non-breakup case with identifying three
\textit{scaling exponents} and find the exponents for the breakup case
dependent on a length scale, which is a \textit{continuous} number. This
dependence is in contrast with (I) our previous studies
\cite{nakazato2018self,nakazato2022air} and (II) the standard critical
phenomena such as the ferromagnetic transition. This is because (I) in our
previous studies the corresponding scaling exponents were found to be always
the same value even if some length scales are changed and (II) in the standard
critical phenomena the exponents usually depend on not continuous but
\textit{discrete} numbers, such as the number of components of the order
parameter $n$, representing the symmetry, and dimensionality $d$. Since, in
critical phenomena, such dependence is known to categorize universality
classes, the dependence on a continuous number in the preset hydrodynamic
analog indicates the existence of an uncountably infinite number of
universality classes, as in some exotic cases known in critical phenomena. The
deep analogy revealed in the present study suggests a promising avenue for the
future study of self-similarity in PDEs by exploring many examples in natural
phenomena and industrial processes, in which confined geometries are relevant.

\begin{figure*}[ptb]
\centering
\includegraphics[width=11.4cm]{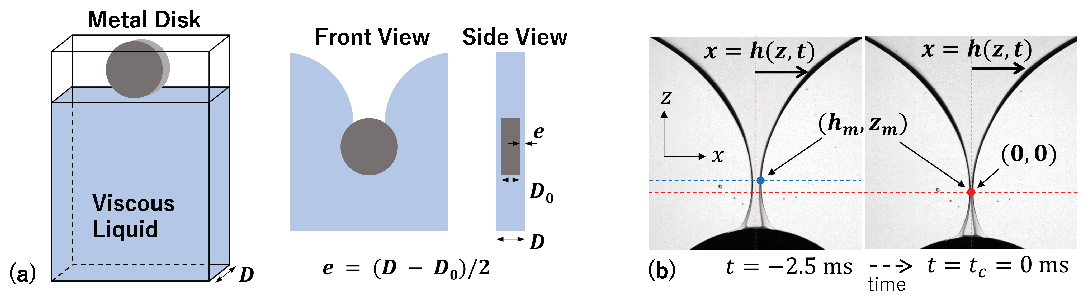}\caption{(a) Experimental setup. A
metal disk of thickness $D_{0}$ ($=1$ mm) and radius $R$ ($10~$to $12.5$ mm)
falls in the cell of thickness $D$ ($2$ to $6$ mm) filled with a viscous
liquid of kinematic viscosity $\nu$ (100 to 1000 cS). The disk entrains air
into the liquid, which finally detaches from the disk. The difference between
$D$ and $D_{0}$ defines the liquid film thickness $e$. (b) Snapshots just
before and at breakup illustrating the setting of axes, in the case with
topology change for $(R,D_{0},e,\nu)=(10,3,0.5,100)$ in mm or cS.}%
\label{Fig1}%
\end{figure*}

\begin{figure*}[ptb]
\centering
\includegraphics[width=11.4cm]{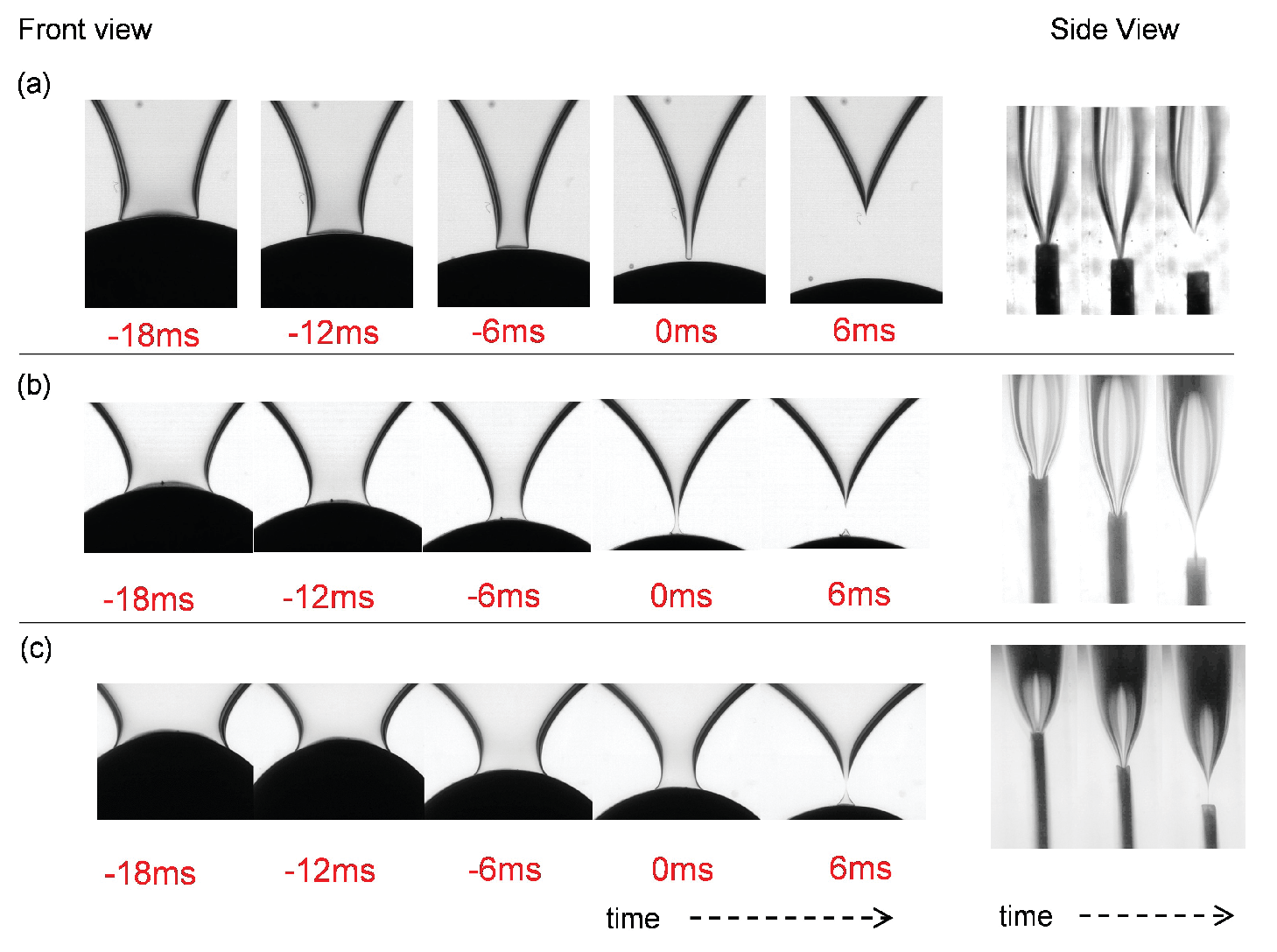} \caption{Snapshots of entrainment of
air by a disk into liquid, leading to detachment of air from the disk for
$R=10$ mm, $D_{0}=1$ mm, $\nu=100$ cS. The liquid film thickness $e$ are 1,
1.5, and 2 mm, respectively, in (a) to (c). The time label 0 ms corresponds to
$t=t_{d}$ defined in the text. In the three rightmost front-view photos reveal
that a small bubble remains on the solid surface in (b) and (c) while such a
bubble cannot be seen in (a): \textit{the breakup transition point} lies
between $e=$ $1$ and 1.5 mm. Side-view shots near the detachment (0 ms) are
separated by 6 ms, but are not synchronized with the front-view shots (a set
of side-view snapshots are obtained from an experiment performed on a day
different from the day on which the corresponding front-view snapshots but
conducted for the same parameters).}%
\label{Fig2}%
\end{figure*}

\section{Results}

\subsection{Experimental\label{S2}}

In Fig.~\ref{Fig1} (a), we show our experimental setup with explanation. The
ranges of characteristic lengths, the radius and thickness of disk and the
cell thickness, are as follows: $R=10-12.5$ mm, $D_{0}=1$ mm, $D=2-6$ mm. The
cell width and height are much larger than the length scales $R,D_{0}$, and
$D$ (typically 9 and 12 cm, respectively). We use polydimethylsiloxane (PDMS)
for viscous liquid, where the range of kinematic viscosity $\nu=\eta/\rho$ is
100-1000 cS. The density $\rho$ and the surface tension $\gamma$ are slightly
depending on viscosity $\eta$ ($\rho\simeq$ $0.97$ g/cm$^{3}$ and
$\gamma\simeq20$ mN/m). The density $\rho_{s}$ of the metal disk (SUS430) is
7.7 g/cm$^{3}$ with the density difference $\Delta\rho=\rho_{s}-\rho$. The
cell is fabricated with acrylic plates of thickness 5 mm, using acrylic
spacers whose thickness defines the cell thickness $D$.

To obtain reproducible results, we set a gate at the top of cell by gluing a
pair acrylic plates of thickness very close to $e=(D-D_{0})/2$, one for the
back surface of the front cell plate and the other for the front surface of
the back cell plate, to make the gap at the gate close to the disk thickness
$D_{0}$. This gate helps to make the thickness of two liquid films between the
surfaces of the disk and cell precisely equal to $e$ (see the Side View in
Fig.~\ref{Fig1} (a)). We fall the disk so that the initial speed of the disk
is zero, i.e., the bottom of the disk is in contact with the interface at zero
velocity at the entry. The disk surface is coated with a very thin layer of
the same liquid as the one in the cell, by once dipping the disk into the
liquid and then removing the liquid well with liquid-absorbing paper, to
guarantee the zero static contact angle. We record the shape change of the
air-liquid interface with a high-speed camera (FASTCAM Mini UX 100, Photron)
with a lens (Micro NIKKOR 60 mm f2.8G ED, Nikon). The range of frame per
second (fps) is 1000-2000. The images are analyzed with Image J and self-made
Python codes.

In the previous studies on the pre-detachment dynamics
\cite{nakazato2018self,nakazato2022air}, the sheet-forming with breakup and
the corn-forming regime without breakup are typically observed for
$D_{0}>\kappa^{-1}$ and $D_{0}<\kappa^{-1}$ at $e=0.5$ mm respectively, where
$\kappa^{-1}=\sqrt{\gamma/(\rho g)}\simeq1.8$ (In the study on the
post-breakup dynamics \cite{Yoshino}, the case of the sheet-forming regime
with breakup was studied). In the present study on the pre-detachment
dynamics, we set $D_{0}$ to a fixed value 1 mm and change $e$ to find the
transition from the breakup to the non-breakup in the corn-forming regime.

\subsection{Shape of the air-liquid interface $h(z,t)$}

In Fig.~\ref{Fig1} (b), we explain the setting of coordinates in the present
study (we set $z_{c}=0$, so that $z-z_{c}=z$ and $z_{m}-z_{c}=z_{m}$). The
air-liquid interface formed by air entrained by the disk is seen as the dark
curve with a finite width. The width reflects the three-dimensional characters
of the interface: the interface is concave or convex. The region where the
width is thin, the interface is almost flat. We track the right and left inner
edges of the thick curve and describe them by $x=h(z,t)$ and $-h(z,t)$,
respectively, except for the case of $e=0.5$ mm, for which we track instead
the outer edges. We determine whether we track the inner or outer edge such
that we can have a better collapse of shapes at different time after
rescaling, which we discuss below. We call the minimum of the function
$h(z,t)$ with respect to $z$ "the constriction point" at which $(x,z)=(h_{m}%
(t),z_{m}(t))$, i.e., $h_{m}(t)=h(z_{m}(t),t)$.

\subsection{Dynamics dependent on the film thickness $e$}

In Fig.~\ref{Fig2}, we show snapshots of air-detachment near the air-solid
contact for $e=1.0$, 1.5, and 2 mm, with the other parameters, $R$, $D_{0}$,
and $\nu$ fixed (we have tried to analyze the case of $e=2.5$ mm, only to fail
because the inner edge was not clear). In three rightmost front-view photos
(labeled as 6 ms), detachment without topological change is observed for $e=1$
mm but that with topological change (i.e., breakup) for $e=1.5$ and $2$ mm: a
small bubble remains on the solid surface in (b) and (c) while such a bubble
cannot be seen in (a). In other words, "the transition point between the
breakup and non-breakup" or \textit{the breakup transition point} lies between
$e=$ $1$ and 1.5 mm.

The origin "0 ms" of the time label ($t-t_{d}$) given in ms in Fig.~\ref{Fig2}
and figures below for the discussion of self-similar dynamics is set for the
snapshots either just before or after the detachment and thus can be slightly
deviated from the real detaching time $t_{d}^{\prime}$ with an error, which
should be at most 0.5 ms when images are recorded at 2000 fps.

In the previous studies \cite{nakazato2018self} and \cite{nakazato2022air},
the cases of $D_{0}\gtrsim\kappa^{-1}$ and $D_{0}\lesssim\kappa^{-1}$ with
$\kappa^{-1}=\sqrt{\gamma/(\rho g)}\simeq1.8$ mm are distinguished by the
terms sheet-forming and corn-forming detachments, and the former and latter
cases are respectively regarded as the cases with and without topology
transition and with and without appearance of the constriction point. However,
in the present case, as seen in Fig.~\ref{Fig2}, both the cases with and
without topology transition are observed for $D_{0}\lesssim\kappa^{-1}$ while
the tip form seen from the side-view snapshots looks like a corn rather than a
sheet and the constriction point, which is a sign of the breakup, can appear.
In other words, we find that the transition from breakup to non-breakup can
occur within the corn-forming detachment.

\subsection{Dynamics of characteristic length scales}

\begin{figure*}[ptb]
\includegraphics[width=17.8cm]{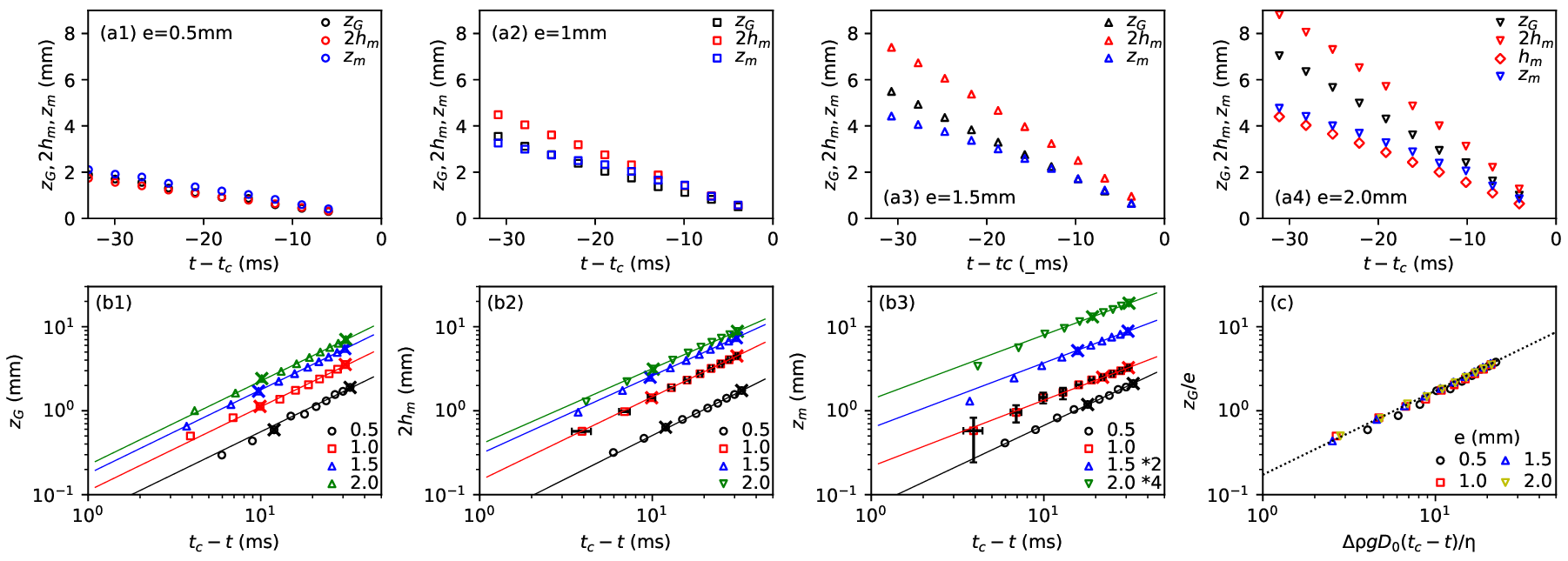} \caption{(a) Plots of $h_{m}$,
$z_{m}$, and $z_{G}$ as a function of $t-t_{c}$ for $e=0.5$, 1, 1.5 and 2 mm
at $D_{0}=1$ mm and $R=10$ mm, where $t_{c}$ is the critical time precisely
defined in the text. In (a4), $h_{m}$ is comparable to $z_{m}$. (b) Plots in
(a), regrouped and plotted on a log-log scale. Solid lines are obtained by
fitting (using the data between the two crosses) with a function with a
function $y=ax$ for $z_{G}$ and $y=ax^{b}$ for 2$h_{m}$ and $z_{m}$ (see the
text for details). In (b3), the quantities $2z_{m}(t)$ and $4z_{m}(t)$ instead
of $z_{m}(t)$ are plotted for $e=1.5$ and 2.0 mm, to avoid overlap with
$z_{m}(t)$ for $e=1.0$ mm. (c) Renormalized plot of $z_{G}(t)$, where the
dotted line represents Eq.~(\ref{eq1a}) with $k=0.254$. As discussed below,
the slope in (b2) [(b3)] determines the exponent $\beta$ [$\Delta$].}%
\label{Fig3}%
\end{figure*}

In Fig.~\ref{Fig3} (a1) to (a4), we show the dynamics of three characteristic
length scales, $2h_{m}(t)$, $z_{m}(t)$, and $z_{G}(t)$, for different film
thickness $e$. The quantity $z_{G}(t)$ is the position of the center of
gravity of the disk measured from that at the critical time $t=t_{c}$, which
slightly differs from $t_{d}$ [$z_{G}(t_{c})=0$ by definition]. The critical
time $t_{c}$ is determined by extrapolation as the time when $h_{m}$ becomes
zero. The critical space-time point is represented as $(x,z,t)=(x_{c}%
,z_{c},t_{c})$ with $x_{c}=h_{m}(t_{c})$ and $z_{c}=z_{m}(t_{c})$.

Plot (a1) shows that three length scales are identical, i.e.,
\begin{equation}
2h_{m}(t)=z_{m}(t)=z_{G}(t), \label{eq1c}%
\end{equation}
at $e=0.5$ mm as also observed in different parameter ranges
\cite{nakazato2018self,nakazato2022air}. In the present case, however, as
shown in plots (a2) to (a4), they grow into different length scales as $e$ increases.

We regroup and show plots in (a) on a log-log scale in (b), which suggest the
existence of scaling laws for these length scales, although the range of
scaling is rather limited to nearly over, or slightly less than, one order of
magnitude. In (b1), since we expect $z_{G}$ linearly scales with $t_{c}-t$ as
discussed below, we fit the data by a linear function $y=ax$ with a fitting
parameter $a$ in the region indicated by a pair of cross marks. The fitting
line with slope one on the log-log scale thus obtained are shown in (b1). In
(b2) and (b3), we fit the data by $y=ax^{b}$ with two fitting parameters $a$
and $b$ in the region indicated by a pair of cross marks. The fitting line
with slope $b$ thus obtained are shown in (b2) and (b3). The log-log plots in
(b1) to (b3) suggest, in addition to the linear scaling for $z_{G}$, the
following scaling laws for $h_{m}$ and $z_{m}$:
\begin{equation}
h_{m}=c_{1}t^{\prime\beta}\text{ and }z_{m}=c_{2}t^{\prime\Delta} \label{eq9}%
\end{equation}
with a time label $t^{\prime}$ (which is positive at times before the critical
time $t=t_{c}$):
\begin{equation}
t^{\prime}=t_{c}-t.
\end{equation}
The exponents $\beta$ and $\Delta$ thus obtained are summarized in
Tab.~\ref{T1}, which shows a systematic dependence of the exponents on $e$.
This is in contrast with the previous studies
\cite{nakazato2018self,nakazato2022air}, where $\beta$ and $\Delta$ are always
one even if $D_{0}$ and $R$ are changed.

\begin{table}[ptb]
\caption{Exponent $\beta$ and $\Delta$ determined from Fig.~\ref{Fig3} by
fitting the data with the functions, $h_{m}(t)\sim(t_{c}-t)^{\beta}$ and
$z_{m}(t)\sim(t_{c}-t)^{\Delta}$, respectively. See the text for the
determination of the exponent $\delta$.}%
\label{T1}
\begin{center}
$%
\begin{array}
[c]{|c||c||c||c||c|}\hline
e\text{ (mm)} & \beta & \Delta & \delta & \Delta/(\beta\delta)\\\hline\hline
0.5 & 1.0\pm0.0 & 0.97\pm0.06 & 1.01\pm0.01 & 0.97\pm0.06\\\hline
1.0 & 1.0\pm0.0 & 0.79\pm0.02 & 0.64\pm0.00 & 1.2\pm0.03\\\hline
1.5 & 0.93\pm0.00 & 0.78\pm0.01 & 0.61\pm0.00 & 1.4\pm0.02\\\hline
2.0 & 0.90\pm0.01 & 0.77\pm0.03 & 0.59\pm0.00 & 1.5\pm0.06\\\hline
\end{array}
$
\end{center}
\end{table}

We carefully selected the range of fitting indicated by a pair of cross marks
for $h_{m}$ and $z_{m}$ to avoid including the data with large error bars and
to use only the data with small error bars, by noting that (i) the error
coming from the time resolution tends to become large near $t=t_{c}$ and (ii)
the error tends to be large in $z_{m}$ compared with $h_{m}$. These issues are
visualized by the error bars shown for the data at $e=1.0$ mm in (b2) and
(b3): The error bars in the horizontal axis are based on the time resolution
$\pm0.5$ ms at 2000 fps; The error bars in the vertical axis are on the
spacial resolution corresponding to 1 pixel of the images for $h_{m}$ but
possibly corresponding to several pixels for $z_{m}$ because of the elongation
explained below. In accordance with Remark (i) and (ii), we can confirm the
following in (b2) and (b3): the error bars in the horizontal axis is visible
near $t=t_{c}$, while the error bars in the vertical axis are not visible for
$h_{m}$ but significant for small $z_{m}$. The reason for large errors in
$z_{m}$ is as follows. The constriction region tends to elongate near the
breakup and the thin thread of air thus formed detaches from the disk when
$h_{m}$ is not zero (see Fig.~\ref{FigS1} in Appendix \ref{ap1}). Because of
this elongation of the constriction region the determination of $z_{m}(t)$
tends to become difficult near breakup, which is reflected in larger standard
deviation for $\Delta$ (compared with $\beta$) in Tab.~\ref{T1}.

As for the fitting of $z_{G}$, Fig.~\ref{Fig3} (c) further justifies the
linear fitting demonstrated in Fig.~\ref{Fig3} (b1), supporting the following
scaling law for $z_{G}(t)$ in agreement with the result established in a
similar parameter range \cite{nakazato2022air}:%
\begin{equation}
z_{G}(t)=v_{G}t^{\prime}\text{ for }t<t_{c}, \label{eq1}%
\end{equation}
with a characteristic velocity scale
\begin{equation}
v_{G}=k\Delta\rho gD_{0}e/\eta\label{eq1a}%
\end{equation}
with the gravitational acceleration $g$. Note that the agreement between the
data and the dotted line shown in (c) is demonstrated with using the value,
$k=0.254$, which is the value obtained in the previous study
\cite{nakazato2022air}.

\subsection{Self-similarity in the interface shape dynamics}

\begin{figure*}[ptb]
\includegraphics[width=17.8cm]{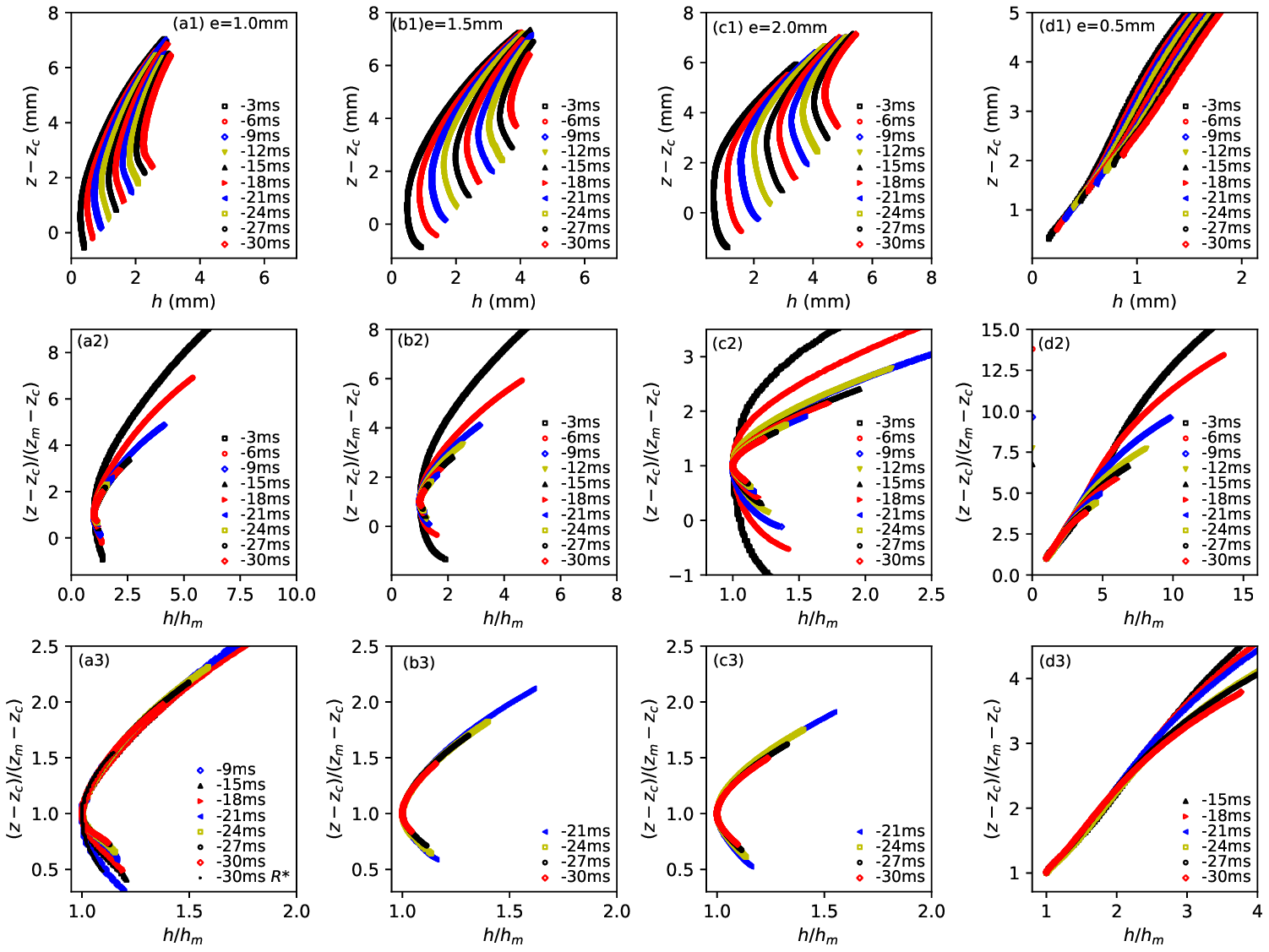} \caption{Shape change for $e=1$, 1.5,
2.0, and 0.5 mm for $(R,D_{0},\nu)=(10,1,100)$ in mm or cS. Since the right
and left interfaces are the mirror image of the other, only the right
interface after averaging is shown. The error bars in capturing the interface
are less than the size of markers. The time development of interfacial shape
before [after] rescaling is respectively shown in (a1) [(a2) and (a3)] for
$e=1$ mm. We see a better collapse for times not too close to $t=t_{c}$
(before $-9$ ms) [see (a3)]. The corresponding plots for $e=$1.5, 2.0, and 0.5
mm are shown respectively in (b) to (d). The data labeled $R^{\ast}$ in (a3)
are obtained for different parameter set $(R,D_{0},\nu)=(12.5,1,100)$, which
collapse well on the master curve [the good collapse seems limited for the
upper branch (of our focus), i.e., the shape above the constriction point].}%
\label{Fig4}%
\end{figure*}

In Fig.~\ref{Fig4} (a1), we show the interfacial shape before rescaling at
different times for $e=1$ mm. Interface shapes after rescaling are reasonably
collapsed well on a master curve, especially near the constriction point, as
seen in (a2). The collapse is significant for shapes before $t-t_{d}=-9$ ms as
shown in (a3).

Similar collapses are shown in (b) and (c) for $e=1.5$ and 2.0 mm,
respectively. As $e$ increases the constriction region tends to elongate near
breakup as announced, which can be confirmed by comparing (b1) with (c1) (see
also Fig.~\ref{FigS1} in Appendix \ref{ap1}). This elongation is related to
the above-mentioned \textit{breakup transition point}: topology does change
for $e=1.5$ and 2 mm with accompanied by breakup but it does not for $e=1.0$
mm. It seems that topological change (i.e., breakup) tends to elongate the
constriction region near $t=t_{d}$. Because of this elongation, duration of a
good collapse tends to be limited as $e$ increases [see (a2) to (c3) and (a3)
to (c3)].

Reproducibility and universality are also demonstrated in Fig.~\ref{Fig4}. The
data labeled $R^{\ast}$ in (a3) imply that the corresponding data are obtained
for different $R$ with the other parameters fixed. In Fig.~\ref{Fig4} (a3),
even these data obtained for different $R$ collapse well on the master curve,
which demonstrates the master curve is independent from (i.e., universal for
change in) $R$ for a fixed $e$, indirectly representing a good reproducibility
of the experiment.

In Fig.~\ref{Fig4} (d1), we show the interfacial shape for $e=0.5$ mm, which
is fundamentally different from cases of $e$ larger than 0.5 mm. In this case,
as mentioned above, we need to track the outer edge instead of inner edge of
the dark interface to observe collapse of the renormalized shape seen in (d2)
[and more clearly in (d3)]. In addition, the constriction point does not
appear [the point $(h_{m},$ $z_{m})$ is defined as the right contact point (at
$y=0$) of the interface with the disk] but the tip forms a corn. However, the
corn may not be axisymmetric but elliptic as reported in our previous study
\cite{nakazato2022air}.

The self-similarity demonstrated in Fig.~\ref{Fig4} can be represented as%
\begin{equation}
h(t,z)=h_{m}(t)\Gamma(z/z_{m}(t))\equiv h_{m}(t)\Gamma(\xi), \label{eq2}%
\end{equation}
where the master curve is represented by the function $\Gamma(\xi)$ with
$\xi=z/z_{m}(t)$. Figure~\ref{Fig4} further suggests that the master curve is
dependent on $e$. In fact, comparing the master curve in (a3) to (c3) with
(d3), the shape difference is evident. Even among (a3) to (c3), we can regard
the shape difference: as $e$ increases, the shape tends to be compressed in
the vertical direction. In Fig.~\ref{Fig4}, we also notice that the master
curve seems to be well described by
\begin{equation}
\Gamma(\xi)=1+c(\xi-1)^{1/\delta}, \label{eq5b}%
\end{equation}
which scales as $\Gamma(\xi)\sim\xi^{1/\delta}$ for $\xi\gg1$. This form and
the exponent $\delta$ will be quantified below.

\begin{figure*}[ptb]
\includegraphics[width=17.8cm]{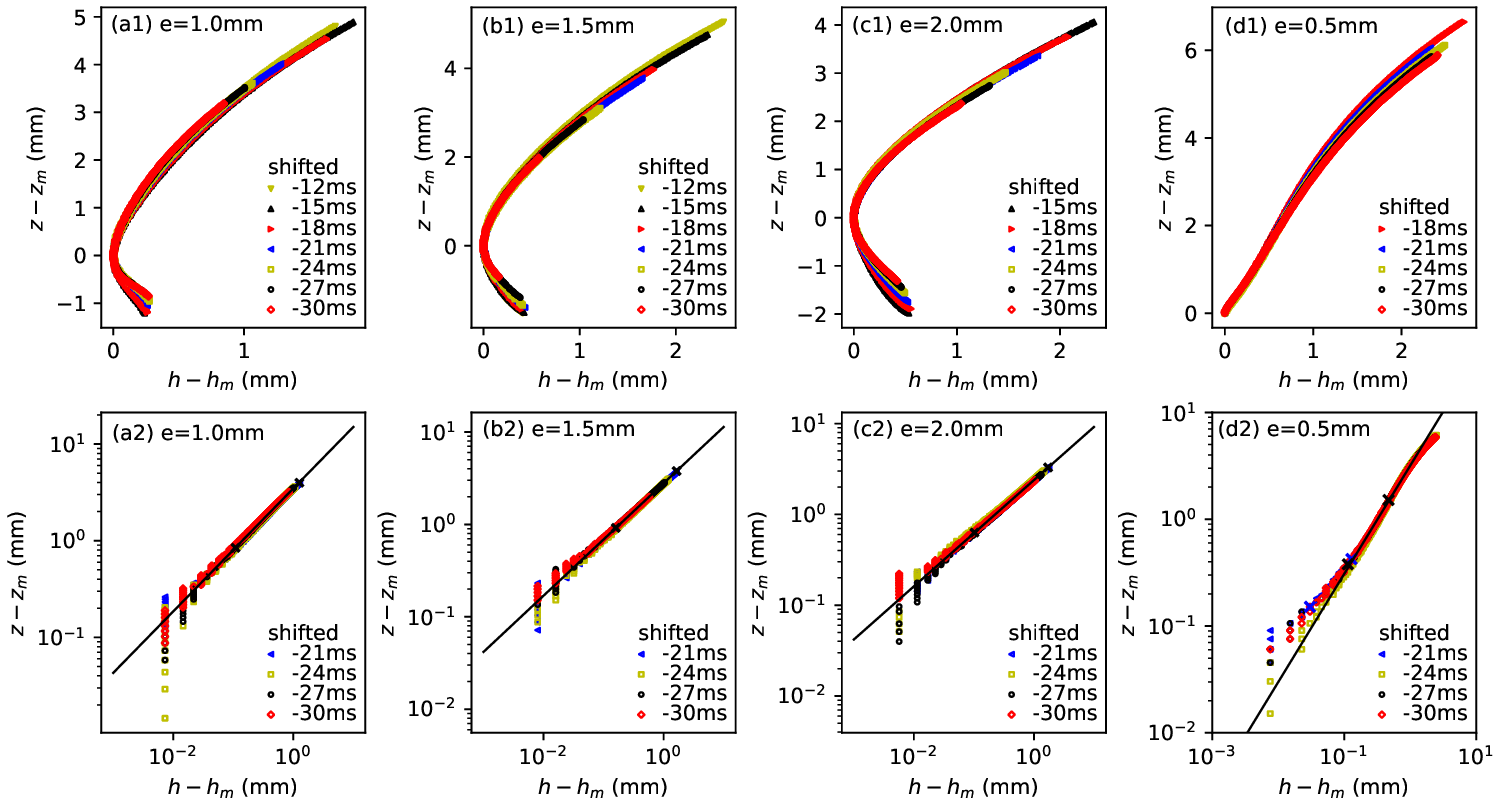}\caption{Translated shape functions
for $e=1.0,1.5,2.0$, and 0.5 mm on linear scales in the period of good
collapse [(a1) to (d1)] and on log-log scales for $-21$ to $-30$ ms with a
fitting line obtained by fitting the data in the region between the two black
crosses for $-21$ ms [(a2) to (d2)]. For all cases, $(R,D_{0},\nu)=(10,1,100)$
in mm or cS.}%
\label{Fig5}%
\end{figure*}

\subsection{Identical shapes translating in time}

We show in Fig.~\ref{Fig5} (a1) to (d1) that the shapes are exactly identical
for a given parameter set in the period not too close to $t=t_{c}$. They are
just translating in space: when $z-z_{m}$ is plotted as a function of
$h-h_{m}$, all the curves collapse onto a master curve. These master curves
can be reasonably well described by the relation,
\begin{equation}
h-h_{m}=c_{0}(z-z_{m})^{1/\delta}\equiv c_{0}\widetilde{z}^{1/\delta},
\label{eq4b}%
\end{equation}
as shown in Fig. \ref{Fig5} (a2) to (d2). The exponent $\delta$ is obtained by
fitting the data in a range of small error bars shown by a pair of black cross
marks in the plots (a2) to (d2), where the fitting lines thus obtained are
shown. Numerical values of $\delta$ thus obtained are summarized in
Tab.~\ref{T1}, which shows $\delta$ is systematically dependent on $e$ (In the
previous study \cite{nakazato2022air}, the data corresponding at $e=0.5$ mm
are fit by a curve with $\delta=0.75$, which is consistent with the present
data: if we fit the data in the region between the pair of blue cross marks,
we obtain $\delta=0.75$, which shows the present fitting focus on a region of
small errors).

The exponent $\delta$ in Eq.~(\ref{eq4b}) can be identified with $\delta$ in
Eq.~(\ref{eq5b}) because the former equation can be cast into the following
form: $h/h_{m}=1+c^{\prime}(z_{m}^{1/\delta}/h_{m})(\xi-1)^{1/\delta}$ with
$\xi=z/z_{m}$, where the collapse of the function at different times observed
in Fig. \ref{Fig4} means that the coefficient $z_{m}^{1/\delta}/h_{m}$ is a
constant,%
\begin{equation}
h_{m}\sim z_{m}^{1/\delta}. \label{eq5c}%
\end{equation}
In other words, Eq.~(\ref{eq4b}) reduces to Eq.~(\ref{eq5b}) and thus $\delta$
in the both equations can be identified with each other.

Equation (\ref{eq5c}), together with Eq.~(\ref{eq2}), leads to the following
relation:%
\begin{equation}
\Delta=\beta\delta. \label{eq17}%
\end{equation}
As seen in the column $\Delta/(\beta\delta)$ in Tab.~\ref{T1}, Eq.~(\ref{eq17}%
) is reasonably well satisfied at $e=1.0$ mm, and less satisfied as $e$
increases. Correspondingly, the spacial region of good collapse becomes more
limited as $e$ increases.

\subsection{Closed equation describing the shape}

The translating interface observed in Fig. \ref{Fig5} can be described by the
expression, $x=h_{m}+f(z-z_{m})$, from which we obtain $\partial h/\partial
t=\dot{h}_{m}-f^{\prime}\dot{z}_{m}$ and $\partial h/\partial z=f^{\prime}$.
These relations lead
\begin{equation}
\frac{\partial h}{\partial t}+\dot{z}_{m}\frac{\partial h}{\partial z}=\dot
{h}_{m}\text{.} \label{eq11}%
\end{equation}
This equation with Eq. (\ref{eq9}) is a closed equation for the shape dynamics.

This equation can be derived from another expression of the translating
interface: the velocity of points on the surface
\begin{equation}
(u_{s}(t,z),v_{s}(t,z))\equiv(u(t,x=h(t,z),z),v(t,x=h(t,z),z)) \label{eq10}%
\end{equation}
is independent of spacial coordinates. This expression means that the
spacially constant velocity should be equal to the speed of the constriction
point:
\begin{equation}
(u_{s},v_{s})=(\dot{h}_{m},\dot{z}_{m}), \label{eq8}%
\end{equation}
for which Eq.~(\ref{eq9}) holds. Equation (\ref{eq11}) can be derived as
announced, if Eq. (\ref{eq8}) is combined with the following equation of
motion for the interface $h$ under Eq.~(\ref{eq8}):%

\begin{equation}
\frac{\partial h}{\partial t}+v_{s}\frac{\partial h}{\partial z}=u_{s}.
\label{eq7}%
\end{equation}
This can be derived from the fact that a point on the interface $(x,z)$ with
$x=h(t,z)$ translates with a velocity $(u_{s}(t,z),v_{s}(t,z))$ defined in Eq.
(\ref{eq10}), i.e., the point moves to the point $(x^{\prime},z^{\prime
})=(x+u_{s}dt,z+v_{s}dt)$ with $x^{\prime}=h(t+dt,z^{\prime})$ after $dt$.

\section{ Discussion}

\subsection{Comparison with the bubble breakup in non-confined geometry}

The translation of the interface in time with keeping the same shape observed
in the present study is a property shared by the breakup of a bubble in
non-confined geometry as explained below. Under no confinement, the breakup
occurs with the axisymmetry. Accordingly, the shape can be described in the
cylindrical coordinate $(r,z,\theta)$ by $r=h(t,z)$, where the shape function
$h(t,z)$ is independent of the $\theta$ coordinate from the axisymmetry. When
a high viscosity $\eta$ of the surrounding fluid competes with capillarity
characterized by the surface tension $\gamma$, the dynamics is known to be
described by the following equation (e.g., see Sec.~9.3.1 of
\cite{eggers2015singularities}):%
\begin{equation}
\frac{\partial h(t,z)}{\partial t}=-\gamma/\eta\label{eq20}%
\end{equation}
Since the equation of motion of the interface in Eq.~(\ref{eq7}) is valid also
in this case, Eq.~(\ref{eq20}) means the surface velocity is given as%
\begin{equation}
(u_{s},v_{s})=(-\gamma/\eta,0).
\end{equation}
This means that the shape is just translating in time in the horizontal
direction as in the present case but with no translation in the vertical direction.

\subsection{Dimensional analysis\label{Sec2}}

We can physically understand how the scaling structure in Eq.~(\ref{eq2}) with
Eqs.~(\ref{eq1c}) and (\ref{eq1}) at $e=0.5$ mm emerges in a natural manner.
The key observation is the present problem can be regarded as finding a
solution for Navier-Stokes equation for a viscous liquid, neglecting inertia
and the role of air. To solve the problem we specify the boundary conditions,
which are characterized by length scales $R$, $e$, and $D_{0}$. In addition,
we can give a condition at $t=t_{c}$ to simplify the problem: the tip is
moving at the velocity of the falling disk $v_{G}$. This velocity given in
Eq.~(\ref{eq1a}) is determined by a viscosity-gravity balance for a falling
disk \cite{nakazato2022air}: the viscous dissipation per time $\eta
(v_{G}/e)^{2}R^{2}e$ occurring in the film of thickness $e$ whose volume
scales as $R^{2}e$ balances with the gravitational energy change per time
$\Delta\rho gR^{2}D_{0}v_{G}$. Furthermore, we note that the scaling structure
in Eq.~(\ref{eq2}) with Eqs.~(\ref{eq1c}) and (\ref{eq1}) is characterized by
dimensional parameters $\eta$, $\Delta\rho$ and $g$ alone, except for length
scales $R$, $e$, and $D_{0}$, but is not involved with $\gamma$ and $\rho$,
which again supports the dynamics governed by a viscosity-gravity balance.
Accordingly, dimensionally, we expect%
\begin{equation}
h(t,z)=f(t^{\prime},z,\Delta\rho,\eta,g,R,e,D_{0}) \label{eq3}%
\end{equation}
we have 9 dimensional variable, of which only 6 are independent, since the
dimension of the unit of all the 9 quantities can be derived from the three
fundamental units, kg, m, and s. From the Buckingham $\pi$ theorem
\cite{Lemons_2017}, we expect a relation $\pi_{0}=\Xi(\pi_{1},\pi_{2}%
,\ldots,\pi_{5})$, where $\pi_{i}$'s are 6 independent dimensionless variables
and $\Xi$ is a dimensionless function. We select these dimensionless variables
by focusing on 7 independent length scales $h$, $z$, $e$, $D_{0}$, $R$,
$l=\eta/(\Delta\rho gt^{\prime})$ and $h_{m}\simeq z_{m}\simeq v_{G}t^{\prime
}$, and set $\pi_{0}=h/h_{m}$ and $\pi_{1}=z/z_{m}$. To select the remaining 4
independent variables, we normalize the remaining 4 scales $e$, $D_{0}$, $R$,
and $l$ by $z_{m}$:
\begin{equation}
h=h_{m}\Xi(z/z_{m},e/z_{m},D_{0}/z_{m},R/z_{m},l/z_{m}) \label{eq4}%
\end{equation}
Here, we may expect that near the breakup point where $h_{m}\simeq z_{m}$ is
small so that $e/z_{m}$, $D_{0}/z_{m}$, $R/z_{m}$ and $l/z_{m}\simeq
1/t^{\prime2}$ eventually approach towards infinity. This implies that if the
function $\Xi$ remains finite and non-zero, it becomes independent of
$e/z_{m}$, $D_{0}/z_{m}$, $R/z_{m}$ and $l/z_{m}$ to recover Eq.~(\ref{eq2})
with Eqs. (\ref{eq1c}) and (\ref{eq1}) at $e=0.5$ mm. In this way, we can
naturally understand the self-similar structure at $e=0.5$ mm based on a
dimensional argument.

In the case of $e=1.0$ to 2.0 mm, in order to justify a similar argument,
different from the case of $e=0.5$ mm, we have yet to know the parameter
dependence of the coefficients $c_{1}$ and $c_{2}$ in Eq.~(\ref{eq9}). Such a
study would be interesting especially in terms of \textit{universality} and
\textit{separation of scales}. For example, if Eq.~(\ref{eq4}) is still valid
even for $e=1.0$ to 2.0 mm, we may understand \textit{universality} of
$\Gamma$ for change in $R$ but not in $e$ indicated in the present experiment:
In the time domain in which the collapse is observed in Fig. \ref{Fig4}, while
$R/z_{m}$ and $l/z_{m}\simeq1/t^{\prime2}$ are large enough compared with the
length scale over which the collapse is observed, $e/z_{m}$ is not large
enough so that the $e$-dependence remains in $\Gamma$. In other words, the
dependence on length scales could disappear if the scale is well separated
from the length scale of collapse, but could remain if not.

This issue of \textit{universality} and \textit{separation of scales} is
explored for the \textit{post-breakup} dynamics \cite{Yoshino}. Similar study
for the present pre-detachment dynamics would be an interesting future
problem, although this requires careful experiments. Note that experiments on
the pre-detachment is technically more challenging because the pre-detachment
dynamics tends to be more sensitively affected by the entry conditions.

We remark that the above tentative argument for $e=1.0$ to 2.0 mm suggests
universality of $\Gamma$ for change in $\eta$. In fact, the master curve in
the previous studies on the pre-detachment dynamics
\cite{nakazato2018self,nakazato2022air} is always independent of $\eta$.

\subsection{Analogy with critical phenomena}

In critical phenomena in thermodynamic transitions, various thermodynamic
quantities exhibit power laws near the critical point. Typical example is the
ferromagnetic transition: the macroscopic magnetization $M(T,H)$ at zero
magnetic field $H=0$ observed at low temperatures $T$ disappears at a critical
temperature $T=T_{c}$, exhibiting a power law: $M(T,H=0)\simeq\Delta T^{\beta
}$ with $\Delta T=\left\vert T-T_{c}\right\vert $. Here and hereafter, for
simplicity, $T$ and $H$ are dimensionless temperature and magnetic field,
respectively. The magnetization is \textit{an order parameter }of critical
phenomena, which is non-zero below $T_{c}$ and is zero above $T_{c}$ at $H=0$.

The power law is observed for different quantities. At $H=0$, the specific
heat $C$ and the susceptibility $\chi$ behave as $C\simeq\Delta T^{-\alpha}$
and $\chi\simeq\Delta T^{-\gamma}$, while the magnetization $M$ behaves as
$M\simeq H^{1/\delta}$ at $T=T_{c}$. The exponents such as $\alpha$ and
$\beta$ are called the critical exponents and their values are governed by
dimensionality $d$ and symmetry characterized by the number of components $n$
of the vector representing the order parameter. In other words, the exponents
do not depend on some details such as the strength of the microscopic
interaction and the structure of lattice. As a result, various materials
characterized by the same $n$ and $d$ share the same critical exponents,
forming \textit{a universality class}. Namely, the same set of critical
exponents universally describes critical behaviors such as $C\simeq\Delta
T^{-\alpha}$ and $\chi\simeq\Delta T^{-\gamma}$ of different materials, if
they belong to the same universality class although some quantities such as
the critical temperature $T_{c}$ are material-dependent. Furthermore, the
critical exponents satisfy \textit{scaling relations} such as $\alpha
+2\beta+\gamma=2$ and such relations are universally satisfied by the critical
exponents of any universality classes. Such remarkable features can be
understood based on the scaling hypothesis as explained in Appendix \ref{ap2},
and the scaling hypothesis naturally emerges in the renormalization group
theory \cite{Cardy,Goldenfeld}.

The scaling hypothesis in the magnetic case can be expressed as
\begin{equation}
M(T,H)\simeq\Delta T^{\beta}\Psi(H/\Delta T^{\Delta})\text{,} \label{eq2b}%
\end{equation}
which bears a strong resemblance with the self-similar structure in
Eq.~(\ref{eq2}) with Eq.~(\ref{eq9}). Time $t$ ($=t-t_{c}$) and position $z$
($=z-z_{c}$) correspond to temperature $T$ and external field $H$,
respectively, for "the order parameter $h(t,z)$," a quantity which is non-zero
before $t_{c}$ but is zero after $t_{c}$ at $z=0$: the pre-detachment dynamics
of our focus corresponds to the temperature range below $T_{c}$. The master
curve $\Gamma(\xi)$ corresponds to the function $\Psi(\xi)$, which is called
the scaling function and defines the critical exponent $\delta$ through
$\Psi(\xi)\simeq\xi^{1/\delta}$ for $\xi\gg1$ (and thus $\delta$ in
Eq.~(\ref{eq5b}) corresponds to the critical exponent $\delta$). The scaling
exponents $\beta$ and $\Delta$ in Eq.~(\ref{eq9}) correspond to $\beta$ and
$\Delta$ in Eq.~(\ref{eq2b}), where the exponents $\beta$, $\Delta$, and
$\delta$ in critical phenomena satisfy Eq.~(\ref{eq17}) as shown in Appendix
\ref{ap2}.

In other words, the critical exponents $\beta$, $\Delta$, and $\delta$ in the
present case are the ones summarized in Tab.~\ref{T1}. This table suggests
that these three exponents are independent, where Eq.~(\ref{eq17}) is more
violated as $e$ increases, and that that the exponents are monotonic functions
of the thickness $e$. In addition, we may expect from the above arguments on
the universality of $\Gamma(\xi)$, which defines one of the exponents $\delta
$, the exponents are universally shared by the systems with different $R$ (and
$\eta$) for a given $e$ (at $D_{0}=1$ mm). Namely, the present universality
class may be determined by \textit{a continuous parameter} $e$. This is in
contrast with typical critical phenomena, in which universality class is
determined by \textit{discrete parameters} such as $n$ and $d$, although
examples of critical exponents depending on continuous parameters are known
when marginal operators play a role, which includes non-linear diffusion
equations \cite{goldenfeld1990anomalous}.

\section{Conclusion}

We investigate the pre-detachment dynamics of a bubble surrounded by a viscous
liquid in a confined geometry focusing on the dependence on the lubricating
film thickness $e$ for a fixed $D_{0}$ ($=1.0$ mm). The present results are in
contrast with our previous studies on the pre-detachment dynamics
\cite{nakazato2018self,nakazato2022air} in that (I) we add a third regime, the
corn-forming regime with breakup, to the previously found two regimes, the
sheet-forming regime with breakup and the corn-forming regime without breakup,
and that (II) we find the scaling exponents characterizing the self-similarity
are dependent on a length. We elucidate similarity between the present
singular dynamics and critical phenomena, showing that the counterpart of the
critical exponents are dependent on a continuous variable $e$ and could be
shared by a universality class formed by the systems with different $R$ (and
$\eta$) for a fixed $e$ (at $D_{0}=1.0$ mm). The dependence on the continuous
variable suggest the existence of an uncountably infinite number of
universality classes. The physical origin of the dependence on $e$ but not on
$R$ could be understood from separation of scales, as explored in the study of
the post-breakup \cite{Yoshino}.

The analogy between the singular dynamics and critical phenomena elucidated in
the present study underscores the importance of exploring symmetry in the
future study of the singular dynamics by exploiting confined geometries for
the following reasons. (1) In critical phenomena, which has propelled the
development of the modern physics from soft and hard condensed matter to
non-equilibrium systems and active matter
\cite{de1979scaling,livi2017nonequilibrium,altland2023condensed,tailleur2022active}%
, dimensionality and symmetry have played a vital role in its development. (2)
The singular dynamics is widely observed in nature and a variety of confined
geometries are increasingly important in industrial applications and natural
phenomena, from microfluidics \cite{StoneStroockAjdari2004,anna2016},
petroleum industry \cite{HeleShawPetroleum2010}, to geology
\cite{parmigiani2016bubble}. In fact, it is recently shown that the analogy
with critical phenomena can be deepened further by developing a
renormalization group analysis for the breakup of a bubble in the non-confined
geometry \cite{Okumura2025RG}, which will be further explored elsewhere for
the present case.

\begin{figure*}[ptb]
\centering
\includegraphics[width=11.4cm]{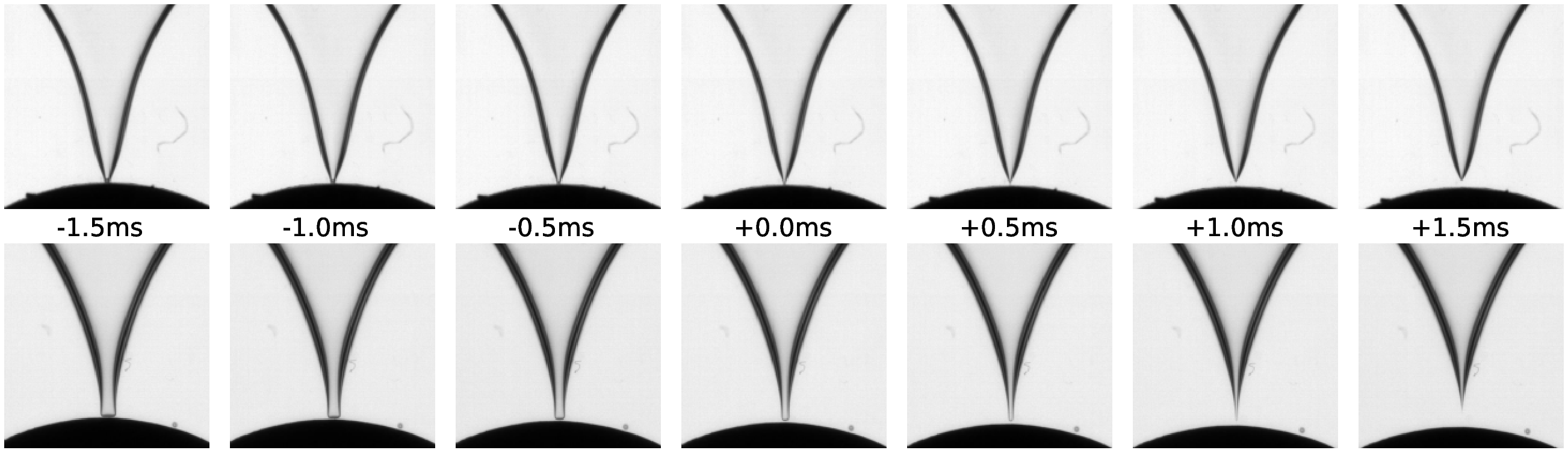}\caption{Snapshots obtained at 2000
fps near the detachment point for $e=0.5$ mm (top) and $e=1.0$ mm (bottom) at
$t-t_{d}=-1.5$ to $+1.5$ ms, respectively. The snapshots at the center
correspond to $t-t_{d}=-0$ ms, where "the order parameter" $h_{m}$ is not
precisely zero.}%
\label{FigS1}%
\end{figure*}

\begin{acknowledgments}
This work was supported by JSPS KAKENHI Grant Number JP19H01859 and JP24K00596.
\end{acknowledgments}


\appendix

\section{Dynamics close to the breakup time\label{ap1}}

We compare in Fig.~\ref{FigS1} the dynamics close to $t=t_{c}$, which shows
qualitative differences for $e=0.5$ and 1.0 mm. In the former, sharp tip is
formed, which detaches from the disk. In the latter, the constriction region
elongates near $t=t_{c}$, forming a thin thread of air, which detaches from
the disk when $h_{m}$ is not zero.

\section{Scaling relations in critical phenomena\label{ap2}}

We summarize here the basic knowledge on critical phenomena, focusing on the
temperature range below $T_{c}$ ($T<T_{c}$). Excluding the exponent describing
the spacial correlation, there are 4 exponents, $\alpha,\beta,\gamma$, and
$\delta$, which are defined in the text as (c1) $C\simeq\Delta T^{-\alpha}$,
(c2) $M\simeq\Delta T^{\beta}$, and (c3) $\chi\simeq\Delta T^{-\gamma}$ at
$H=0$, together with (c4) $M\simeq H^{1/\delta}$ at $\Delta T=0$. However,
among the four exponents, only two are independent. This results from the
following thermodynamic relations, which are proven to be true by virtue of
the renormalization group: The free energy $F(T,H)$ possesses the scaling
structure (t1) $F(T,H)\simeq\Delta T^{2-\alpha}\Phi(H/\Delta T^{\Delta})$ near
the critical temperature with a scaling function $\Phi(x)$, where
thermodynamic quantities are derived from $F(T,H)$ as (t2) $C\simeq
\partial^{2}F/\partial T^{2}$, (t3) $M\simeq\partial F/\partial H$,\ and (t4)
$\chi\simeq\partial^{2}F/\partial H^{2}$. With these thermodynamic relations
(t1) to (t4), the definitions of the critical exponents (c1) to (c4) result in
the relations (a) $\Delta T^{-\alpha}\Phi(H/\Delta T^{\Delta})\simeq\Delta
T^{-\alpha}$, (b) $\Delta T^{2-\alpha-\Delta}\Phi^{\prime}(H/\Delta T^{\Delta
})\simeq\Delta T^{\beta}$ (i.e., another scaling function $\Psi(x)$ introduced
above in Eq.~(\ref{eq2b}) is expressed as $\Psi=\Phi^{\prime}$), (c) $\Delta
T^{2-\alpha-2\Delta}\Phi^{\prime\prime}(H/\Delta T^{\Delta})\simeq\Delta
T^{-\gamma}$ at $H=0$ and (d) $\Delta T^{2-\alpha-\Delta}\Phi^{\prime
}(H/\Delta T^{\Delta})\simeq H^{1/\delta}$ at $\Delta T=0$. By requiring,
based on consistency, that $\Phi(0)$, $\Phi^{\prime}(0)$, and $\Phi
^{\prime\prime}(0)$ are non-zero finite value, while relation (a) simply means
consistency, relations (b) and (c) lead to the following two independent
relations, (A) $2-\alpha-\Delta=\beta$ and (B) $\beta-\Delta=-\gamma$, which
result in the scaling relation $\alpha+2\beta+\gamma=2$ discussed in the text.
Furthermore, relation (d) concludes, again based on consistency, the
asymptotic behavior $\Psi(x)\simeq x^{1/\delta}$ for large $x$ with
Eq.~(\ref{eq17}), \textit{if Eq.~(\ref{eq2b}) is valid even for }$x=H/\Delta
T^{\Delta}$\textit{ is large}. In this way, we can prove three independent
relations (A), (B) and Eq.~(\ref{eq17}) for 5 exponents, $\alpha,\beta$,
$\gamma$, $\delta$, and $\Delta$, which proves that there are only 2
independent exponents. However, Eq. (\ref{eq2}), which corresponds to
Eq.~(\ref{eq2b}) in the magnetic case, holds only for relatively small
$\xi=z/z_{m}(t)$, as seen in Fig.~\ref{Fig4}. This suggests breaking of
Eq.~(\ref{eq17}), which allows the existence of three independent exponents,
as implied in Tab.~\ref{T1}.



\end{document}